\documentclass[letterpaper,twocolumn,10pt]{article}
\usepackage{usenix-2020-09}
\usepackage{booktabs}
\usepackage[]{babel}
\usepackage{paralist, tabularx}
\newcommand{\ourmethod}{\texttt{ScaAR}\xspace}
\usepackage{makecell}
\usepackage{circledsteps}

\begin{document}

\date{}

\title{\Large \bf  Real-world Edge Neural Network Implementations Leak Private Interactions \\ Through Physical Side Channel}

\author{
{\rm Zhuoran Liu}\\
Radboud University \\
z.liu@cs.ru.nl
\and
{\rm Senna van Hoek}\\
Radboud University\\
senna.vanhoek@ru.nl
\and
{\rm P\'eter Horv\'ath}\\
Radboud University\\
peter.horvath@ru.nl
\and
{\rm Dirk Lauret}\\
Radboud University\\
dirk.lauret@ru.nl
\and
{\rm Xiaoyun Xu}\\
Radboud University\\
xiaoyun.xu@ru.nl
\and
{\rm Lejla Batina}\\
Radboud University\\
lejla@cs.ru.nl
} 

\maketitle

\begin{abstract}
Neural networks have become a fundamental component of numerous practical applications, and their implementations, which are often accelerated by hardware, are integrated into all types of real-world \emph{physical devices}.
User interactions with neural networks on hardware accelerators are commonly considered privacy-sensitive. 
Substantial efforts have been made to uncover vulnerabilities and enhance privacy protection at the level of machine learning algorithms, including membership inference attacks, differential privacy, and federated learning.
However, neural networks are ultimately implemented and deployed on physical devices, and current research pays comparatively less attention to privacy protection at the implementation level.
In this paper, we introduce a generic physical side-channel attack, \ourmethod, that extracts user interactions with neural networks by leveraging \emph{electromagnetic (EM) emissions} of physical devices. 
Our proposed attack is \emph{implementation-agnostic}, meaning it does not require the adversary to possess detailed knowledge of the hardware or software implementations, thanks to the capabilities of deep learning-based side-channel analysis (DLSCA).
Experimental results demonstrate that, through the EM side channel, \ourmethod can effectively extract the class label of user interactions with neural classifiers, including inputs and outputs, on the AMD-Xilinx MPSoC ZCU104 FPGA and Raspberry Pi 3 B.
In addition, for the first time, we provide side-channel analysis on edge Large Language Model (LLM) implementations on the Raspberry Pi 5, showing that EM side channel leaks interaction data, and different LLM tokens can be distinguishable from the EM traces.

\end{abstract}

\section{Introduction}
\label{sec:introduction}

Neural Networks are becoming increasingly integrated into everyday applications, from medical analytics~\cite{tchito2021biomedical} to conversational AI~\cite{openai2025chatgpt}.
This transformative impact extends to autonomous vehicles~\cite{xiao2023nnvehicles}, financial analysis~\cite{hu2021surveyfinancialanalysis}, personalization~\cite{zhang2019deep}, and scientific research~\cite{jumper2021highly}, driving technological advancements and addressing real-world challenges.
All neural networks are implemented and executed on physical devices, including specialized accelerators such as GPUs~\cite{krizhevsky2012imagenet} and TPUs~\cite{jouppi2017datacenter}, and resourced-constrained real-world edge devices such as micro-controllers~\cite{saha2022machine} or FPGAs~\cite{ma2017optimizing}. 
This dependence serves as the cornerstone of neural network performance and efficiency, facilitating everyday applications.

As neural networks evolve and diversify their applications, the demands on hardware have become increasingly pronounced.
Meeting these application demands requires the fast and tailored development of advanced hardware accelerators and the integration of neural network algorithms~\cite{mohaidat2024survey}.
Additionally, given the highly sensitive nature of many neural network applications, such as those in healthcare, finance, and personal assistance, robust privacy protection and security measures at the hardware level are becoming increasingly critical~\cite{Oliynyk2023:IKnowWhatYouTrainedLastSummer, horvath2024sok}. 
To this end, hardware robustness serves as the fundamental element in realizing neural networks' full capabilities while making sure that the data being processed are secure and private. 

Current research on the security and privacy of neural network implementations mainly focuses on model extraction and input recovery~\cite{horvath2023barracuda, Gongye2023:SCA-DPU, Maia2021:HearShapeNN, Thu2023:YouOnlyGetOneShot, Batina2019:PosterInput, huegle2023power2picture, horvath2024sok}.
Model extraction is a well-studied research direction in the past few years. 
Attackers aim to reconstruct the neural network's architecture~\cite{Hu2020:Deepsniffer, Chmielewski2021:OnReverseEngineering}, hyperparameters~\cite{Maia2021:HearShapeNN, Joud2023:SPAonNetworks, Liang2022:Clairvoyance}, and parameters~\cite{Batina2019:CSI-NN, Gongye2023:SCA-DPU, horvath2023barracuda} by exploiting information leaked during its operation, such as timing, and power consumption.
With a different goal of compromising user privacy, input recovery attacks are getting more attention.  
After extracting the model parameters,  well-established DPA attacks can be directly mounted to extract explicit input data by power and EM analysis~\cite{Batina2019:PosterInput, Maji2021:LeakyNets}. 
Another line of input recovery attack exploit specific neural network implementation, including floating-point multiplication~\cite{Dong2019:FloatingPointMult} 
, zero-skipping (i.e., zero-value model )~\cite{Thu2023:YouOnlyGetOneShot, Wei2018:IKnowWhatYouSee}.

\begin{figure}[ht]
    \centering
    \includegraphics[width=\linewidth]{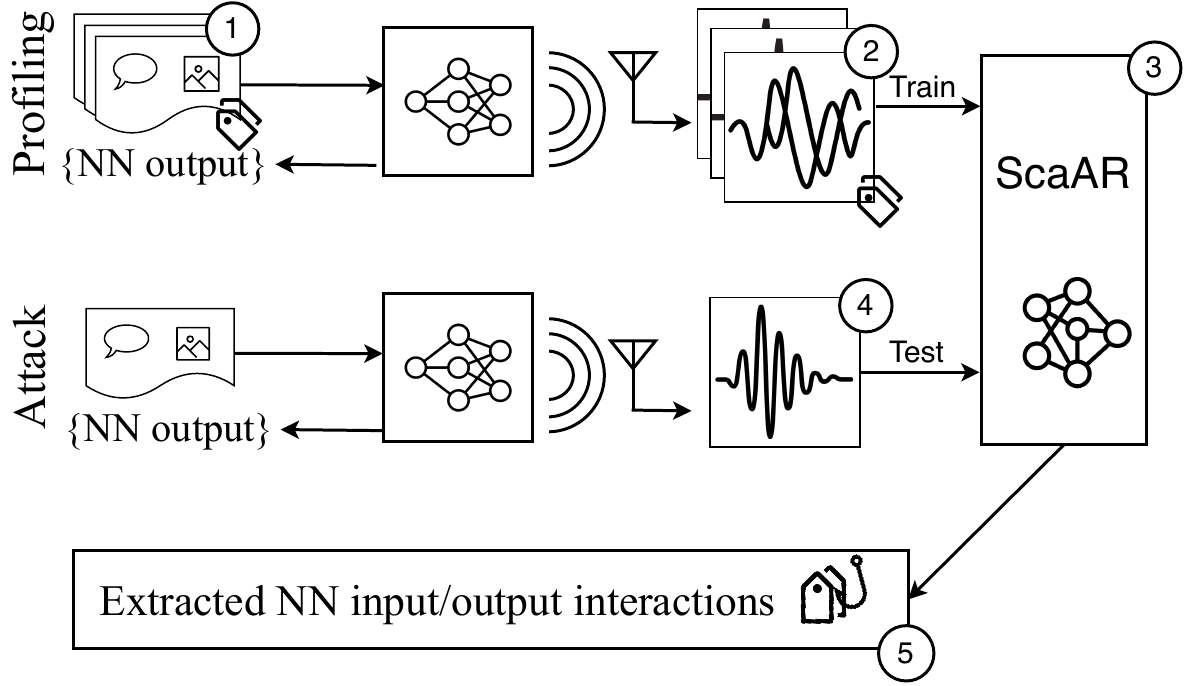}
    \caption{Diagram illustrating the working mechanism of \ourmethod.
    The top row represents the \emph{profiling stage}, where the adversary collect EM traces (\textcircled{\footnotesize2})from a physical device that runs neural network implementations, e.g., classifier, taking user interactions as inputs (\textcircled{\footnotesize1}). Combining the input data annotations, e.g., class labels, and the collected traces, \ourmethod (\textcircled{\footnotesize3}
) is trained and deployed for the attack. The middle and bottom rows represent the \emph{attack phase}. EM traces (\textcircled{\footnotesize4}) are collected from a target physical device. By feeding collected traces to \ourmethod, the adversary can predict the private attributes of the user interactions (\textcircled{\footnotesize5}).}
    \label{fig:teaser}
\end{figure}

Even though current input recovery attacks study the potential threats of side-channel attacks on compromising the privacy of users interacting with the neural network system, the practical threat is still limited.
First, strong assumptions are made in the current attacks. 
Time-to-digital converter (TDC) circuits are used as a probing module~\cite{Moini2021:RemoteBNN}, and weights of the victim neural network implementation are known~\cite{Batina2019:PosterInput}. 
Such attacks provide valuable information for evaluating the worst-case vulnerabilities, but their assumptions may not hold or can be easily mitigated in practice. 
For example, FPGA accelerators can be physically isolated for different users in cloud computing,
and weight extraction  incurs significant computational overhead~\cite{ Gongye2023:SCA-DPU, horvath2023barracuda}. 

More critically, most attacks are \emph{design-informed}, meaning that the adversary is aware of the design details of the hardware or the software implementations. 
Exploiting design-informed vulnerabilities facilitates leakage localization and improves the attack strategy, but it also limits the ability to generalize the attack to unseen threat scenarios. 
In addition, related to the accelerator design, the hard threshold is used as a common method to extract interaction information~\cite{Wei2018:IKnowWhatYouSee, Dong2019:FloatingPointMult, Thu2023:YouOnlyGetOneShot, Moini2021:RemoteBNN, Maji2021:LeakyNets}, which is commonly implementation-specific and further limits the applicability of the proposed attacks.

In this paper, we systematically revise the current input recovery attacks, considering the complete threat life cycle, and refine the threat model to make it better fit the real-world threat scenario (See Section~\ref{sec:ThreatModel}). 
In particular, we argue that the exact extraction of the user interactions with neural network implementations is always unnecessary, and knowing the critical private attribute of the interaction data could already meet the adversarial incentive (See Section~\ref{sec:notnecessary}). 
This refined threat model realizes a more practical threat scenario with a fine-grained adversarial incentive.

Following this refined threat model, we propose a novel Side-channel attack for private Attribute extRaction (\ourmethod) that predicts the attributes of input and output from electromagnetic emissions when the user interacts with the neural network running on physical devices. 
Figure~\ref{fig:teaser} illustrates the workflow of \ourmethod. 
\ourmethod is implementation-agnostic profiling attack~\cite{chari2003template} that is not sensitive to trace \emph{averaging}, and it does not need clock-cycle-level alignment.
Specifically, in the profiling phase, as shown in the top row of Figure~\ref{fig:teaser}, the adversary collects EM traces from the target device, annotates the traces, and builds a machine learning model that predicts the private attribute of the user interactions. 
In the attack phase, the trained model can be used to predict the private attribute of user interactions based on the collected traces of the target device when running the inference task. 
Our experiments demonstrate that the methodology behind \ourmethod generalizes to different physical devices, different data sets, and different victim model architectures (See Section~\ref{sec:experiments}), which suggests that our attack is implementation-agnostic and generalizable.

In sum, we make the following main contributions:

\begin{itemize}
    \item After revisiting the threat model and assumptions of previous input recovery attacks, we show that the generalization of previous attacks is limited, where most attacks are \emph{design-informed} and \emph{implementation-specific}.  
    
    \item We propose a generic attack, \ourmethod, with a refined methodology that is applicable on different physical devices without relying on specific knowledge of certain hardware devices or assumptions. 
    Our attack also does not depend on reverse engineering the implementation before mounting the attack. 
    \item Experimental results demonstrate that \ourmethod is effective in extracting input and output class labels when targeting neural network classifiers on two real-world physical devices, AMD Xilinx MPSoC ZCU104 FPGA and Raspberry Pi 3 B. 
    We also provide further analysis of the proposed attack, which further confirms the utility of \ourmethod.
    We also anticipate that \ourmethod methodology can be leveraged in building secure machine learning hardware.
    \item We also, for the first time, provide a pilot side-channel analysis on edge Large Language Model (LLM) implementations. 
    Our experiments demonstrate EM side channel leaks the number of output tokens, and different LLM tokens can be distinguishable solely from their EM traces.

\end{itemize}

\section{Background}
\label{sec:background}

In this section, we provide the general background of neural networks and physical side-channel analysis.

\subsection{Neural Networks}
Neural networks are algorithms that approximate functions to solve a given task. 
In a supervised setting, such as image classification, the NN represents a function that maps input images to their corresponding classes. 
In the context of neural network implementations, there are three primary attack targets: the input, the architecture, and the parameters.

The inputs to the neural network can have different forms such as images and text, depending on the particular task. 
The architecture and the parameters define the function that maps the inputs to their corresponding outputs. 
The architecture is also dependent on the particular task, such as CNNs for image classification~\cite{lecun1995convolutional} and Transformers~\cite{vaswani2017attention} for natural language processing. 
Given an objective, appropriate parameters are learned during the training of the neural network by feeding input output pairs to the network and modifying the parameters~\cite{lecun1989backpropagation}. 
After training, the trained model can be deployed to perform the given task and make \emph{inference} on new, unseen input data.
In this work, we aim to recover the input attributes by collecting EM traces on physical devices that run neural network implementations. 

\subsection{Side-Channel Analysis}

Side-Channel Analysis (SCA) can be applied to obtain information about the executed operations and processed data by an electronic device that leaks unintentional information via various \emph{side channels}, such as timing~\cite{dhem2000practical}, power~\cite{Kocher1998Power} or electromagnetic emanations~\cite{gandolfi2001electromagnetic}.

\noindent\textbf{Power analysis.} Power analysis exploits the dependency between the dynamic power consumption of an electronic device and the executed operations and used data~\cite{Kocher1998Power}. 
Simple Power Analysis (SPA) and Differential Power Analysis (DPA) are forms of SCA where an adversary can extract secret information, such as the used cryptographic algorithm and key~\cite{Kocher1998Power}, by measuring the power consumption of an electronic device. 
SPA typically requires only one or few traces while DPA requires significantly more, as it is typically performed using statistical methods.

\noindent \textbf{Electromagnetic analysis.} Similarly to power analysis, the electromagnetic (EM) emanations of an electronic device can be measured to extract secret information.
In addition, \emph{Simple Electromagnetic Analysis (SEMA)} and \emph{Differential Electromagnetic Analysis (CEMA)} work in the same way as SPA and DPA in power analysis.
In practice, EM analysis has been used to not just break cryptographic implementations, but also to eavesdrop on display units~\cite{kuhn1998soft, gandolfi2001electromagnetic,van1985electromagnetic,elibol2012realistic,hongxin2009recognition, liu2020screen}.

\noindent \textbf{Profiled attacks.} Profiled attacks require the adversary to be in possession of the target device or an unprotected device that is identical to the target device~\cite{chari2002template}.
The attacker then profiles the device to model the signal and noise expected of the target device. 
These attacks are powerful alternatives for adversaries when they cannot obtain a large number of traces (for the same secret) from the target device, as required by DPA/DEMA.  
Profiled attacks include template attacks ~\cite{chari2002template,archambeau2006template} as well as Deep Learning Side-Channel Analysis (DLSCA) ~\cite{gilmore2015neural,lerman2015machine,maghrebi2016breaking,bartkewitz2012efficient,lerman2014power, picek2023sok}.
In this paper, we use DLSCA to extract interaction attributes from raw EM traces. 

\section{Related Work}
\label{sec:relatedwork}
In this section, we cover the relevant literature most closely related to our work. 
Section~\ref{sec:relatedwork:modelextraction} first introduces general model extraction attacks explaining algorithm-level attack without exploiting side channels.
Moving to a real-world scenario, we discuss physical side-channel model extraction attacks.
Section~\ref{sec:relatedwork:inputrec} covers interaction recovery attacks that extract user interactions, especially inputs, with the neural network. 
Section~\ref{sec:relatedworkMLpriv} covers algorithm-level machine learning privacy attacks with a focus on LLMs.
Section~\ref{sec:relatedworddiscuss} discusses common assumptions made in interaction recovery attacks. 

\subsection{Model Extraction}
\label{sec:relatedwork:modelextraction}

\noindent \textbf{Algorithm-level Model Extraction.}
Model extraction attacks have been well studied on the algorithm-level, where the adversary could query the 
target model and build models based on the query-output pairs from the model feedback feedback~\cite{Oliynyk2023:IKnowWhatYouTrainedLastSummer}.
Query-based extraction involves sending carefully crafted inputs to a target model and observing its outputs, such as class labels or confidence scores. 
By using input and query interchangeably,
an attacker can build a surrogate model that mimics the functionality of the target~\cite{Tramer2016:StealingMLModels, Orekondy2019:KnockoffNets}. 
This method is particularly effective in black-box settings where the model's internal architecture and parameters are not accessible. 
Techniques like adaptive querying~\cite{karmakar2024marich} can further improve the attack by focusing on inputs that yield the most informative outputs, thus reducing the number of queries required for an accurate replica. 
Query-based extraction is often used in scenarios where the target model is deployed via APIs, as these typically allow unrestricted or minimally monitored access to predictions~\cite{Oliynyk2023:IKnowWhatYouTrainedLastSummer}.

\noindent \textbf{Physical Side-Channel Model Extraction.}
Physical side-channel attacks offer an alternative approach to extracting models from their implementations. Early side-channel model extraction attacks follow methodologies commonly used for analyzing cryptographic circuits. 
Techniques such as simple power or electromagnetic (EM) analysis can reveal the architecture and even the hyperparameters of target implementations~\cite{Batina2019:CSI-NN}. 
Assuming adversaries have control over the model inputs, correlation-based power or EM analysis can further be employed to extract the model parameters~\cite{Batina2019:CSI-NN, Regazzoni2020:MLHardwareSecurity, Yli2021:ExtractionBNN, Gongye2023:SCA-DPU, horvath2023barracuda}.

To mount a Differential Power Analysis (DPA) attack, the adversary must have a precise understanding of the implementation, which often requires low-level reverse engineering. 
For instance, a simulator called sDPU has been introduced to reverse-engineer DPU activities~\cite{Gongye2023:SCA-DPU}, and binaries from TensorRT have been analyzed to identify parameter representations and relevant partial sums~\cite{horvath2023barracuda}.
It is important to note that DPA-based input recovery attacks heavily rely on the parameters extracted through model extraction, making reverse engineering a necessary step.
In this paper, our work conducts attacks directly on raw traces annotated solely with attribute information. 
This approach enhances the generalizability of our attack, making it more applicable to real-world scenarios.

\subsection{Interaction Recovery}
\label{sec:relatedwork:inputrec}

There are mainly two privacy concerns regarding the user privacy when interacting with a machine learning system.
One concern is whether or not creative content has been used for training deployed neural networks.
Several developers of neural networks are now defending lawsuits against data distributors, e.g., Stability AI that is facing a lawsuit from Getty Images~\cite{reuters_getty_stability_ai_2023}.
As a result, licensing agreements between data owners and neural network developers are being made, e.g., OpenAI has agreements with newspapers for training purposes~\cite{nyt_openai_newscorp_deal_2024}.
Another growing concern is the potential recovery of input data from neural networks, which could expose private information. 
Attacks targeting the recovery of neural network inputs are commonly referred to as \emph{Input Recovery} attacks~\cite{horvath2024sok}. 
For example, security cameras are increasingly deployed with embedded DNN chips for facial recognition, making the input data highly privacy-sensitive.

\begin{table*}[!t]
\caption{Overview of interaction recovery attacks. DL represents deep learning side-channel analysis, SPA represents simple power analysis, DPA represents differential power analysis, and SA represents standard signal analysis.
}
\label{tab:all}
\begin{center}
\begin{small}
\begin{sc}
\resizebox{\textwidth}{!}{
\begin{tabular}{lcclccc}
		\toprule
		
		\textbf{Paper} & \textbf{Platforms} & \textbf{Model Type} & \textbf{Dataset} & \textbf{Side Channel} & \textbf{SCA}\\
\midrule		
		Wei, Lingxiao, et al. (2018)~\cite{Wei2018:IKnowWhatYouSee}                          &
		FPGA               & CNN           & MNIST                                 & Power             & SPA, SA, DPA\\
		
		Batina, et al.(2019)~\cite{Batina2019:PosterInput}                                     &
		              CPU & MLP           & MNIST                                 & EM                & DPA\\
		
		Dong, et al. (2019)~\cite{Dong2019:FloatingPointMult}                                 &CPU & MLP           & MNIST                                 & Power, Timing             & SA\\
		
		Maji, et al. (2021)~\cite{Maji2021:LeakyNets}                                         &CPU         & MLP,CNN, BNN           & MNIST, CIFAR-10, ImageNet             & Power, Timing             & SPA, Profiled Attack\\
		
		Moini, et al. (2021)~\cite{moini2021power}                                            &FPGA& BNN           & MNIST                                 & Power      & SA\\
		
		Huegle, et al. (2023)~\cite{huegle2023power2picture}                                  &
		FPGA & CNN           & MNIST                                 & Power      & Profiled Attack (DL)\\
		
		Thu, et al. (2023)~\cite{Thu2023:YouOnlyGetOneShot}                                   &
		FPGA& BNN           & MNIST                                 & EM                & SA\\
\midrule		\ourmethod (Ours)                                                                             &
		FPGA, CPU & \makecell{MLP, CNN \\ Transformer}           & MNIST, CIFAR-10, ImageNet             & EM                & Profiled Attack (DL) \\
		
		\bottomrule
	\end{tabular}
}
\end{sc}
\end{small}
\end{center}
\end{table*}

One approach towards recovering the input of neural networks, is to use side-channel analysis on the target platform.
SCA can be performed in two different ways either with physical side channels or with non-physical side channels.
Non-physical side channels refer to the side-channels that expose physical quantities of the target network, without having physical access to the target device. For example, practices of non-physical side channels include using loop circuits on shared platforms to obtain power traces~\cite{Moini2021:RemoteBNN}.
For physical SCA, the adversary requires physical access to the device to measure the physical quantities of the target platform, i.e., to measure the power consumption or electromagnetic emanations.
In the following, we only cover physical side-channel interaction recovery attacks.

\noindent \textbf{DPA attacks.}
Most established attack that recovers the inputs of neural networks using physical side channels is DPA~\cite{Batina2019:PosterInput, Maji2021:LeakyNets}.
DPA for input recovery borrows the techniques used in parameter extraction attacks, where power consumption during the convolution step is used to determine candidates for a specific pixel.
The power consumption is then correlated to the possible different intermediate values.
With the candidates for the intermediate values, knowledge of the weights can be leveraged to determine the input values~\cite{Batina2019:PosterInput, Maji2021:LeakyNets}.
Alternatively, a template can be created which maps the possible input values to a power consumption pattern~\cite{Wei2018:IKnowWhatYouSee}.
The observed power consumption can then be correlated to the mapped power consumption patterns to determine the value of the input pixel.

Even though using DPA allows the adversary to explicitly recover the input, there are some limitations that reduce the practical applicability of the approach.
Typically, with DPA, averaged traces are used for correlation to reduce the noise, which is not possible with input recovery, making it harder to distinguish between different signals~\cite{mangard2008power}.
Additionally, performing DPA is considered to be an expensive operation, as for images every pixel requires the intermediate values to be determined through correlation with all possible candidates~\cite{Gongye2023:SCA-DPU, horvath2023barracuda}.
Furthermore, in most cases, DPA requires the parameters of the network to be known, which is mostly infeasible.
The parameters of a network are considered intellectual property (IP) and are typically not publicly known. 
An attacker may decide on extracting the parameters of a network before recovering the input, but the weight extraction assumes that the adversary can control model inputs, which limits the threat scenario.
DPA might not always be the best solution to recover the input, depending on the available resources of the adversary.

Instead of using DPA to recover the input pixel-by-pixel, Generative Adversarial Networks can be used to generate the input image directly from the obtained traces~\cite{huegle2023power2picture}.
The first step towards using GAN for input recovery is to profile and train a generative network to take power traces as input and outputs images that are of the same form as the input.
For training purposes, the adversary needs to be in possession of a profiling device to obtain traces similar to that of the target device.
Then during the exploitation phase, the adversary obtains the power traces and feeds this into the GAN to obtain an approximation of the originally provided input.
State-of-the art GAN approaches towards input recovery have very limited applicability to real-world cases due to data complexity and mode collapse~\cite{thanh2020catastrophic}. 
In addition, it is a prerequisite that the adversary knows the input size of the neural network and the type of images that is being fed to the network.
So even though the input recovery is made less computationally expensive than DPA, the output of GANs might not be the interest of the adversary.

Another approach is to extract the silhouette of the objects in the image which may already reveal privacy-sensitive information about the input~\cite{Wei2018:IKnowWhatYouSee}.
To recover the silhouette of the input, the adversary analyses the individual traces and determines if the consumed power for an intermediate value is higher or lower than some threshold, resulting in a high or low pixel value~\cite{Wei2018:IKnowWhatYouSee, moini2021power}.
Since the multiplications in the convolutional steps are dependent on the intermediate values, larger input values cause the calculations to consume more power and take longer~\cite{Dong2019:FloatingPointMult, Thu2023:YouOnlyGetOneShot, Wei2018:IKnowWhatYouSee}.
Therefore, looking at the consumed power and the time to execute the computations, tells if the input pixel was a big or a small value.
 
Even though the computational footprint of silhouette extraction is significantly less than that of using DPA to recover the entire image, silhouette extraction has only been proven to be effective on black-and-white MNIST input images~\cite{Wei2018:IKnowWhatYouSee}.
Black-and-white images are obviously more suited for silhouette extraction, due to the fact that mapping the pixels to either 0 or 255 does not have much influence.
Colored images may not be well suited for silhouette extraction, since several features in the images could be lost.
Table~\ref{tab:all} summarizes all physical-side channel-based interaction extraction attacks. 
For the reasons mentioned above, most attacks focus on MNIST data set on a physical device. 
Our generic attack, \ourmethod, considers different data sets with various model types on different platforms.

\subsection{LLM Privacy Attacks}
\label{sec:relatedworkMLpriv}

Aside from extracting input images from classifying neural networks, there is rising interest in recovering the input prompts from LLMs.
With the rising popularity of tools like ChatGPT~\cite{achiam2023gpt}, Microsoft's Co-Pilot~\cite{microsoft_copilot}, and Google's Gemini~\cite{team2023gemini}, there is an increasing privacy concern around the processing of the input prompt to those tools.
Prompt extraction~\cite{zhang2024effective} against LLMs aims to convert a prompt to an image, and a prompt to an answer, where side-channel attacks are not .

For LLMs that convert prompts to an answer there are two approaches that are currently being used.
The naive method of extracting the prompt from an answer provided by an LLMs, is simply returning the output to the LLM and asking the network what the input prompt could have been.
Alternatively, the adversary could use a neural network that is trained to link an output to characteristics that are present in the prompt of the target network, e.g., if the text was a business-related prompt, or a code-related prompt.
To this purpose, the adversary needs to first train a network, for which a dataset needs to be created that links labels the characteristics of the input prompt to the output~\cite{sha2024prompt}.
This dataset is then used to train the adversary's network, after which the network is able to predict characteristics of the provided input prompt~\cite{shen2024prompt}.

Although prompt extraction is currently limited to non-side-channel analysis approaches, it has already raised several concerns.
Prompt extraction demonstrates the privacy issues with LLMs, where the adversary can obtain characteristics of the input prompt.
In addition, prompt extraction itself poses a threat towards the intellectual property of prompt engineers~\cite{white2023prompt}.
Physical side-channel-based interaction extraction  has the potential to worsen the situation.

\subsection{Discussions on Physical Side-Channel Interaction Recovery}
\label{sec:relatedworddiscuss}

Table~\ref{tab:Taxonomy} summarized all side-channel interaction recovery attacks.
Current research demonstrates the potential of physical side-channel attacks targeting various physical devices with different implementations on diverse data sets. 
However, the threat scenarios, under which previous attacks are evaluated, are still far from being practical.
Below, we summarize and discuss the most substantial common assumptions made in previous research:

\begin{itemize}
    \item \textbf{Design-informed attacks.} Most previous attacks can either access the assembly (software implementation attacks) or the design (hardware implementation attacks). 
    In the design-informed scenario, the adversary is aware of the exact target operations in both software and hardware designs. 
    For example, preamble in the AXI bus~\cite{Thu2023:YouOnlyGetOneShot} and multiplication difference in the line buffer~\cite{Wei2018:IKnowWhatYouSee, moini2021power} in the hardware design, or zero-skipping operation~\cite{Maji2021:LeakyNets} by inspecting the disassembly code~\cite{Dong2019:FloatingPointMult}. 
    Leveraging the design knowledge relaxes the threat model and facilitates the attack.
    \item \textbf{Hard threshold.} Focusing on the exact operation that leaks the input information, several representative attacks exploit a hard threshold to distinguish background and foreground pixels~\cite{Wei2018:IKnowWhatYouSee} or to distinguish zeros and non-zero pixels~\cite{Dong2019:FloatingPointMult, Maji2021:LeakyNets}. 
    Some threshold selections are guided by the explicit implementation of the accelerator, enabling the adversary to identify the leaky operation and establish the threshold accordingly.
    \item \textbf{Available parameters.}
    DPA attacks assume that the adversary is aware of the model parameters. 
    However, this could be challenging because of both the difficulties of reverse-engineering and the threat model conflict. 
    Considering reverse-engineering, previous works show that this is not easy and even impossible, especially with limited resources~\cite{Gongye2023:SCA-DPU, horvath2023barracuda}.
    About the threat model, parameter extraction always assumes the control of the input, which naturally contradicts with the problem formulation of interaction recovery. 
    It indicates that some attacks may only pose limited threats when parameters are not available. 
    
    \item \textbf{Clock cycle alignment. }
    Per clock cycle alignment is also a crucial part for pixel-level recovery~\cite{Maji2021:LeakyNets}. 
    In practice, it will introduce extra difficulties and may not be feasible. 
    
\end{itemize}

Ideally, a practical interaction extraction attack ought to be implementation-agnostic and generalizable across different physical devices. 
In the remainder of this paper, we aim to clarify the real-world threat and propose a new physical side-channel attribute recovery attack, \ourmethod, tackling those issues.

\section{Problem Formulation}

In this section, we provide a detailed background and propose a new threat model that better resembles real-world threats. 
Section~\ref{sec:notnecessary} discusses why explicit interaction extraction is not necessary and what the adversary is interested in.
Section~\ref{sec:ThreatModel} introduces our threat model that considers real-world scenarios. 
Section~\ref{sec:scaar} covers our attack \ourmethod and introduces our attack methodology and pipeline.

\subsection{Explicit Interaction Extraction is Not Always Necessary}
\label{sec:notnecessary}
Following the strategy of side-channel attacks on cryptographic implementations~\cite{mangard2008power}, 
correlation analysis-based input recovery attacks can extract the exact operands~\cite{Batina2019:PosterInput}.
Such \emph{high-fidelity explicit} attacks extract the exact data that is being processed by the algorithm. 
Explicit extractions are necessary for side-channel analysis of cryptographic implementations since there is only one private key that could indeed break the security of the target device. 
However, explicit interaction extraction is not necessary to harm the owners and users of machine learning and neural network implementations in real-world scenarios.

\noindent \textbf{User privacy.}
User interactions are treated as private~\cite{zhang2023prompts}, and, when authorized, are also part of the core assets of a machine learning company~\cite{fowl2021adversarial}. 
Private information that is included in the user interactions with the machine learning system may not need the exact interaction extraction.
For example, the private attribute of a patient could be whether the patient has a certain kind of disease instead of the complete report~\cite{fredrikson2015model}. 
The adversary only needs to extract information related to the disease to crack the privacy of the patient. 
Same for the users of the LLM-based conversational agents, the sensitive keywords in the conversation between the user and the agent is the most privacy-sensitive part of the interactions.
The adversary may only need to extract such parts to compromise the privacy of the user without extracting the complete conversations. 
In this paper, we aim to extract the private attribute of the interaction instead of extract the explicit interaction data.

\subsection{Threat Model}
\label{sec:ThreatModel}

\subsubsection{Threat Scenario and Adversary Goal}
\label{sec:threatmmodel:attackerscenario}

The primary goal of the adversary is to recover attributes from the input or output of neural network implementations, which may contain privacy-sensitive information.
The input is to be recovered when it is actually being processed in the inference stage of the network, rather than monitoring the buses~\cite{Thu2023:YouOnlyGetOneShot} to the neural network.
There may be various countermeasures in place that prevent the adversary from directly reading the input to the neural network, for confidentiality preservation.
However, the side channels are not subject to any leakage-reducing solutions.
Furthermore, in this scenario, the attacker does not have any knowledge about the parameters, input, and output of the deployed target neural network.

The following use case illustrates the threat scenario:

\emph{Modern phones use neural models to parse and annotate images at rest for retrieval purposes, where sensitive information may be leaked from the output annotations of these neural representation models.
During the preparation phase, the adversary positions an EM antenna near a phone of similar make, while it labels a variety of images to perform profiling. 
For each processed image, the adversary collects the corresponding EM traces and annotates them with attributes of interest to train a machine learning-based attack model. 
During the attack phase, the adversary places a (hidden) antenna near the locked victim device to capture EM traces. 
The trained attack model is then used to extract the attributes of interest from the collected EM traces of the victim device.}

\subsubsection{Adversary Knowledge and Capabilities}

\noindent \textbf{Adversary knowledge.} For the acquisition, we assume that the adversary 
is knowledgeable about how to use the measurement equipment, i.e., antenna, (pre-)amplifier, and oscilloscopes.
For the analysis, we assume that the adversary has enough computational resources to train a DLSCA model to analyze the traces.
The adversary has access to a profiling device sufficiently similar to the target device, which is used to collect training data for the machine learning classifier.
The adversary does not know the parameters of the target neural network implementation. 

\noindent \textbf{Adversary capabilities.} We assume that the adversary has physical vicinity to the hardware on which the inference is taking place, at least once.
Here, the adversary places the probes to monitor the power consumption of the target physical device running the neural network inference.
The context for the attack is a side-channel analysis setup for a passive adversary, i.e., the adversary does not physically touch any parts of the platform ensuring that the victim device is not aware of it's presence.

\begin{table}[t!]
\caption{Capabilities of the adversary in the attack phase. }
\label{tab:Taxonomy}
\begin{center}
\begin{small}
\begin{sc}
\resizebox{\linewidth}{!}{
\begin{tabular}{lcccc}
        \toprule
		\multirow{2}{*}{\textbf{Paper}} & \multicolumn{2}{c}{\textbf{Attacker Knowledge}} & \multirow{2}{*}{\textbf{\makecell{Implementation \\ specific}}}\\
		
		 & \textbf{Arch.} & \textbf{Params.} \\
		\midrule
		
		Wei, et al. (2018)~\cite{Wei2018:IKnowWhatYouSee}                                     & \fullcirc & \emptycirc & \fullcirc\\
		Batina, et al.(2019)~\cite{Batina2019:PosterInput}                                     & \fullcirc & \fullcirc & \fullcirc\\
		Dong, et al. (2019)~\cite{Dong2019:FloatingPointMult}                                  & \emptycirc & \emptycirc & \fullcirc\\
		Maji, et al. (2021)~\cite{Maji2021:LeakyNets}                                          & \fullcirc & \fullcirc & \fullcirc\\
		Moini, et al. (2021)~\cite{moini2021power}                                             & \fullcirc & \emptycirc & \fullcirc\\
		Huegle, et al. (2023)~\cite{huegle2023power2picture}                                  & \emptycirc & \emptycirc & \emptycirc \\
		Thu, et al. (2023)~\cite{Thu2023:YouOnlyGetOneShot}                                    & \fullcirc & \emptycirc & \fullcirc\\

		\midrule
		ScAR (Ours)                                                                            & \emptycirc  & \emptycirc & \emptycirc\\
		
        \bottomrule
	\end{tabular}
}
\end{sc}
\end{small}
\end{center}
\end{table}

Table~\ref{tab:Taxonomy} provides the threat model of \ourmethod and other interaction recovery methods. 
DPA methods need the exact intermediates for the calculation, e.g., correlation~\cite{Batina2019:PosterInput, Maji2021:LeakyNets}, which are powerful but need the exact model parameters. 
In addition, most approaches are implementation-specific, meaning that the proposed attacks may not work on other implementations or need substantial adaptations.
Note that Power2Pic~\cite{huegle2023power2picture} shares the same threat model of our method, and it is also implementation-agonistic. 

However, it needs averaging to eliminate electronic noises, and we also found it hard to re-produce when the dataset is getting more complicated than MNIST due to GAN's mode collapse~\cite{thanh2020catastrophic}. 
\ourmethod is a more generic implementation-agonistic attack since we dont need to explicitly extract input pixels but focusing on the interaction attributes.

\subsection{\ourmethod for Interaction Attribute Extraction}
\label{sec:scaar}
\begin{figure}[t]
    \centering
    \includegraphics[width=1\linewidth]{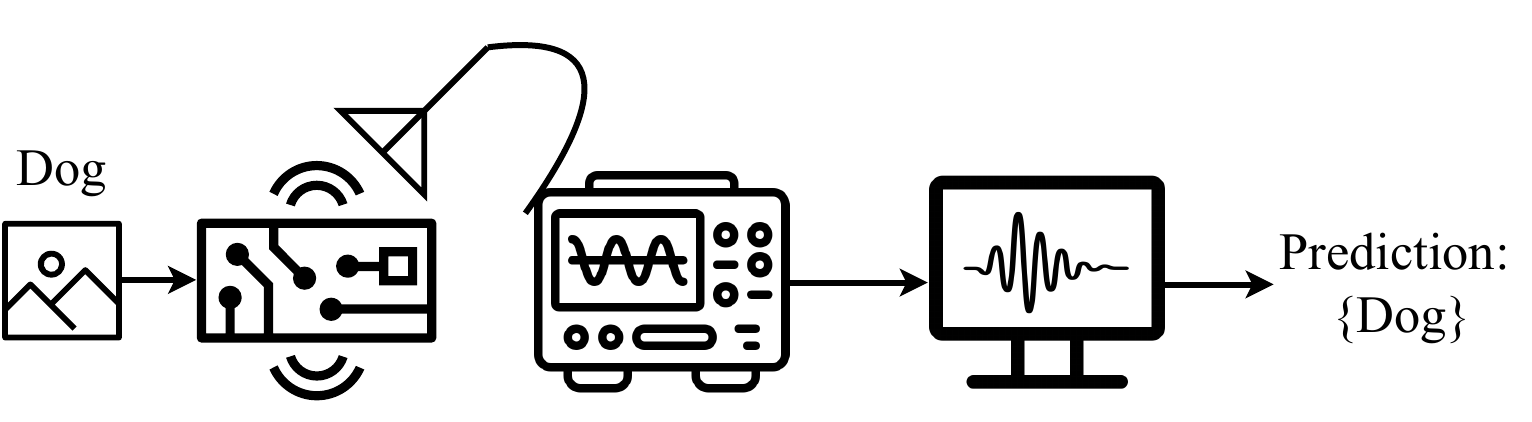}
    \caption{Schematic physical setup for image input attribute extraction. From left to right, a ``Dog'' image is fed into a physical device. The EM emission from the image inference can be collected to predict the attribute ``Dog''. }
    \label{fig:image-input}
\end{figure}

We now introduce \ourmethod in more detail. 
\ourmethod is a profiling side-channel attack that includes a profiling stage and an attack stage. 
The schematic overview of physical setup is shown in Figure~\ref{fig:image-input}.

The device under attack is assumed to be a standard device that runs a neural network, for example, a FPGA with a commercial accelerator IP.
The attacker can only rely on unintentional electromagnetic leakage of the device under attack to reconstruct the user-device interactions.
The leaked electromagnetic signal is first acquired by an oscilloscope and then analyzed on a workstation. 

\noindent \textbf{Profiling phase.} 
First, \ourmethod needs to collect EM traces when running neural network implementations on the device under attack.
Specifically, each collected trace is annotated with labels, after which a \emph{profiling set} is curated. 
Then, the profiling set can be pre-processed including alignment, filtering, or averaging.
We show that \ourmethod can tolerate a certain amount of misalignment and electronic noise (See Section~\ref{sec:misalignment}).
Lastly, a DLSCA \emph{attack model} is trained on the profiling set, and the trained model will be used to mount the attack. 

\noindent \textbf{Attacking phase.} 
EM traces from the target device are collected.
After preprocessing, collected traces are fed into the trained attack model. 
Attack model predicts the sensitive attributes of the input to the device under test, without relying on extra information in the attack phase. 
An illustrative diagram of \ourmethod is shown in Figure~\ref{fig:teaser}, where the top row shows the profiling phase, and the bottom row shows the attack phase.

\section{EM Emissions Reflect User Interactions}
\label{sec:hardware-EM-leak}
In this section, we introduce in detail the experimental setup and provide Simple EM analysis on a a target FPGA. 

\subsection{Experimental Setup}

\subsubsection{Neural Network Implementaion Targets}

We target neural network implementations of Multi-Layer Perception with 5 layers (MLP5), LeNet5~\cite{lecun1998lenet5}, 
vanilla CNN3,
SqueezeNet~\cite{iandola2016squeezenet}, 
and ResNet18~\cite{ResNet}. 
Note that SqueezeNet is used to represent edge-use scenarios where the adversary has a higher chance of being in the vicinity of the target device. 
Table~\ref{tab:acc_all} demonstrates the performance of evaluated neural network implementations. 
MLP and CNNs are evaluated for classification tasks on MNIST~\cite{lecun1998mnist}, CIFAR-10~\cite{CIFAR-10}, and ImageNet-10~\cite{deng2009imagenet} that has 10 classes of images similar to CIFAR-10. 
FPGA models are trained with PyTorch~\cite{paszke2019pytorch} and then quantized and compiled by Vitis-AI 2.5\footnote{\url{https://github.com/Xilinx/Vitis-AI}}.
RPi3B models are trained with Keras~\cite{gulli2017deep} and deployed with TensorFlow Lite~\cite{david2021tensorflow}.
We cover edge Large Language Model (LLM) implementations in Section~\ref{sec:edgellm}.

\begin{table}[t]
\centering
\caption{Classification accuracy (\%) of target neural network implementations on MNIST, CIFAR-10, and ImageNet-10.}
\begin{tabular}{lccc}
\toprule
 & MNIST & CIFAR-10 &ImageNet-10\\
\midrule
MLP    &84.2&54.3& - \\
CNN3    &98.8&-& - \\
LeNet5    &86.4&72.4& - \\
SqueezeNet  &- &91.1&	58.1\\
ResNet18 &90.8&94.1&69.8\\

\bottomrule
\end{tabular}
\label{tab:acc_all}
\end{table}

\begin{figure}[t]
    \centering
    \includegraphics[width=1\linewidth]{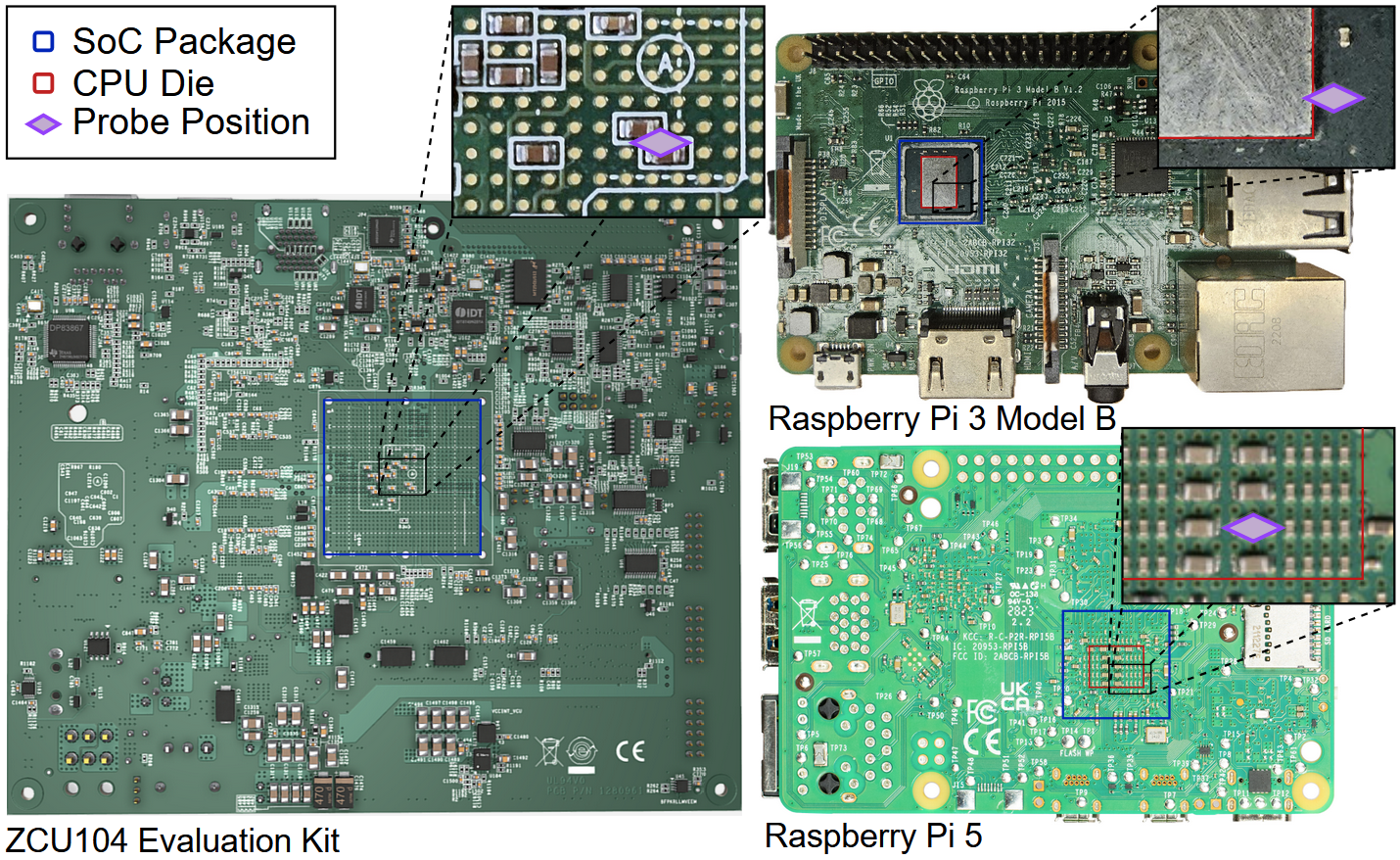}
    \caption{Overview of the target devices. The EM probe position is shown relative to marked areas of interest.}
    \label{fig:devices}
\end{figure}

\begin{table}[t]
\centering
\caption{Overview of targeted physical devices.}
\resizebox{\linewidth}{!}{
\begin{tabular}{lcc}
\toprule
 Device & Target SoC & Clockspeed \\
\midrule
ZCU104 Evaluation Kit   &  XCZU7EV-2FFVC1156 & 100MHz(PL) \\
Raspberry Pi 3 Model B   & BCM2837 B0 &  600 MHz \\
Raspberry Pi 5   & BCM2712 C1 &  300 MHz \\
\bottomrule
\end{tabular}
}
\label{tab:dev_specs}
\end{table}

\subsubsection{Hardware targets}
The ZCU104 Evaluation Kit (ZCU104) features a AMD-Xilinx Zynq UltraScale+ MPSoC, which combines a Armv7-A CPU based Processing System (PS), the with FPGA based Programmable Logic (PL). The PL is configured to act as a neural network accelerator, which will be targeted. The Raspberry Pi 3 model B (RPi3B) and Raspberry Pi 5 (RPi5) (used for LLM in Section~\ref{sec:edgellm}) both feature Broadcom SoC. The CPU architectures of the RPi3B and RPi5 are based on Armv8-A and Armv8.2-A respectively. On the Raspberry Pi boards, the neural network will run on the CPU directly. The configured clockspeeds and specific SoC models are listed in Table \ref{tab:dev_specs}.

We use a passive antenna Langer RF-U 2.5-2 near-field probe probe for all targets. 
The signal from the probe is amplified with a pre-amplifier. 
The signal probed is collected by a Lecroy oscilloscope at a sampling rate of 2.5GS/s. 

\subsubsection{Leakage Localization}
We first need to identify an optimal measurement point on the target 
board, where the EM emissions are highly dependent on the inference task. 
When running a machine learning inference task, we position the probe at various locations on the board and observe the amplitude and shape of the measured EM traces on the oscilloscope. 

\noindent \textbf{FPGA.}
On the ZCU104, we focus on locations that primarily include decoupling capacitors on the opposite side of the main chip and traces entering and exiting the chip.
After the scanning, we observe that a specific capacitor on the back of the board exhibits high amplitude with limited noise, making it optimal for measuring the side-channel EM emissions. 
The placement of this capacitor and the exact probe positioning are shown in Figure~\ref{fig:devices}.

\noindent \textbf{Raspberry Pi.} 
On RPi3B and RPi5 boards, we follow the same procedure of the FPGA. 
The placement of the capacitors and the exact probe positioning are shown in Figure~\ref{fig:devices}.

\subsubsection{EM Trace Acquisition}

\noindent \textbf{Trigger setup.} For FPGA, we use the received signal as trigger, since the PL units and PS units are located in physically distinct areas, and from the PL unit we can get a clear trigger signal from  
The EM signals are highly distinguishable from the electronic noise, and we use the trigger signal for measurements. 
One experiment demonstrates that the EM signal can be collected to mount the attack (See Section~\ref{sec:experiments}) without touching the programmable logic (PL) design.

For the Raspberry Pi, we utilize the onboard GPIO pins and Pi GPIO to configure the trigger. 
All tests are run on Raspberry Pi OS (64-bit, Kernel version: 6.6, Debian version: 12) without real time OS patches. This configuration was chosen over a bare metal implementation or modified kernel to demonstrate the robustness of the attack in the presence of higher level noise sources.
On embedded Linux devices such as the Raspberry Pi, electronic noise is not the only noise source. 
EM emissions from other running applications and scheduled interrupts also contribute to interference in the EM traces of neural network implementations.
However, EM emissions from the neural network inference are distinguishable from electronic noises by SEMA.

\noindent \textbf{Trace collection and preprocessing.}
On the ZCU104, RPi3B, and RPi5, we collect one trace for each inference task on one image during profiling for training unless otherwise specified.
One trace for each image is used for the attack, where the prediction accuracy is reported. 
The collected traces are not aligned before train and test, and we expect the DLSCA attack model to overcome the misalignment. 
Profiling time ranges from 2 hours to 30 hours for different models on each data set.

\subsubsection{EM Trace Analysis by \ourmethod}

We collect 10\,000 traces on the test split of MINST and CIFAR-10 and 13\,000 traces on the validation set of ImageNet, where 90\% traces are used for training and validation (i.e., profilng), and 10\% traces for testing (i.e., attack). 
We treat the class label as the attribute of the data and report prediction accuracy on the test set to resemble the side-channel attack effectiveness. 
Our main DLSCA model is a four-layer one dimensional CNN model~\cite{snyder2018x}.
We also explore other DLSCA models as attack model, including GPAM~\cite{bursztein2024generalized} and 1D-ViT (See Section~\ref{sec:otherdiscussions}).
GPAM is a state-of-the-art transformer-based DLSCA model good at processing de-synchronized traces. 
1D-ViT is reformed by us from the 
original Vision Transformer~\cite{dosovitskiy2020image} to work on one-dimensional input.

\subsection{EM Emissions are Data Dependent} 
\label{sec:EM Emissions are Data Dependent}

In this section, we show that neural network implementations on commercial edge FPGAs are data-dependent. 
We first show that the zero-value model~\cite{mangard2008power} or zero-skipping~\cite{Maji2021:LeakyNets} is clearly visible on Xilinx FPGA.

\begin{figure}[t]
    \centering
    \includegraphics[width=\linewidth]{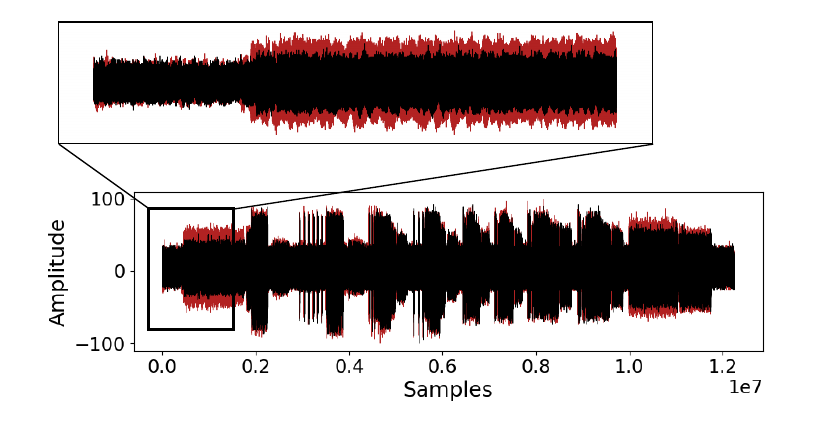}
    \caption{EM traces when running SqueezeNet inference on all-zeros input and regular image input on AMD-Xilinx ZCU104 FPGA. The top row zooms in the first part of the original traces, showing the difference between the two types of inputs (all zeros vs
regular random image input).}
        \label{fig:zcu_sema}

\end{figure}

\noindent \textbf{SEMA on FPGAs}
We visually inspect the data dependency of the EM emission by conducting SEMA on AMD-Xilinx ZCU104 FPGA.
To qualitatively measure the data dependency of the EM emission, we prepare two types of inputs, feed them as inputs of the neural network, and collect EM traces for each input data.  
In particular, images of all zeros and regular random image input are fed as input to the neural network model.
Figure~\ref{fig:zcu_sema} demonstrates that distribution of EM emissions are substantially different for different groups of inputs. 
This finding also confirms that zero-skipping is a possible optimization in commercial FPGA machine learning inference accelerators, which suggests us to differentiate the traces that are introduced by different data.


The SEMA analyses indicate that the EM emissions are dependent on the input data being processed, which hints that fine-grained information can be leaked from the EM traces. 
In the following section, we try to determine more explicit information from EM traces.

\section{Experiments}
\label{sec:experiments}

This section contains main experimental results and analysis. 
Section~\ref{ses:inputattributeextraction} covers the input attribute extraction on different physical devices with different data sets and neural network implementations. 
Section~\ref{sec:lesstrace} explores the minimal amount of traces that makes the attack effective.
Section~\ref{sec:misalignment} covers the generalization of CNN attack models on predicting misaligned traces. 
Section~\ref{sec:additional} provides additional analysis that further interprets \ourmethod.
Section~\ref{sec:otherdiscussions} provides further thoughts
on different DLSCA methods and extraction of an untrained classifier. 
Section~\ref{sec:output} assumes a weaker threat model and evaluates \ourmethod on output extraction.

\subsection{Input Attribute Extraction}
\label{ses:inputattributeextraction}

We first evaluate \ourmethod on extracting attribute of input data. 
We collect and annotate profiling traces on the target devices, based on which we train an attack classifier that predicts the input class. 
In particular, we treat the class of the input data as secret.
For example, in the profiling phase, we collect traces on the target implementation when taking images of dog as input.
In the attack phase, when images of dogs are inferred by the target implementation, we aim to predict the class dog based only on the attack trace.

Table~\ref{tab:main_aligned} demonstrates that \ourmethod is effective in extracting input attributes on the ZCU104 and RPi3B. 
Similar performance is also validated on the RPi3B, which confirms the generalization of \ourmethod.
The ZCU104 is a commercial device whose neural network accelerator IP is confidential, meaning that, even though \ourmethod is not design-informed and different from previous attacks (See Section~\ref{sec:relatedworddiscuss}), so it has a huge potential to pose real-world privacy threats.
\ourmethod is also effective on RPi3B with a running OS.
Even though we reduce the running frequency of the ARM CPU core to fit the bandwidth of our oscilloscope, \ourmethod could successfully extract input attributes from noisy EM traces, which we attribute to the flexibility of the CNN attack model.

\noindent \textbf{Extraction without design modification.} Our experiments are conducted on pre-compiled PYNQ image (version 3.0.1) without any modifications on PS (processing system) and PL (programmable logic) configurations.
The FPGA core and the CPU core are located in physically distinct areas, so we can use local leakage as trigger to collect traces. 

\noindent \textbf{Electronic noise.} 
We collect EM traces without averaging, so we only collect one trace for each user input.
Averaging is critical in eliminating electronic noises, as shown in different attacks~\cite{Dong2019:FloatingPointMult, Moini2021:RemoteBNN}, including the latest input recovery attack~\cite{huegle2023power2picture}.
\ourmethod shows promising results when dealing with traces without averaging, which we believe is the benefit of using deep learning based side-channel analysis.

\begin{table}[t]
\centering
\caption{Input attribute extraction accuracy (\%) on MNIST, CIFAR-10, and ImageNet-10 traces.}
\resizebox{\linewidth}{!}{

\begin{tabular}{lcccc}
\toprule
 Device&Implementation& MNIST & CIFAR-10 &IMN-10\\
\midrule
\multirow{5}{*}{ZCU104}&MLP    & 45.6   &54.7    &-\\
&CNN3 &80.0 &-&-\\
&LeNet5    & 74.5   & -  &-\\
&SqueezeNet  &-    & -  &64.5 \\
&ResNet18 &92.8& 89.4& -\\
\midrule
RPi3B&LeNet5  &-    &96.3   &-\\

\bottomrule
\end{tabular}
}
\label{tab:main_aligned}
\end{table}

\subsection{Extraction with Less Traces}
\label{sec:lesstrace}

\begin{figure}
    \centering
    \includegraphics[width=0.75\linewidth]{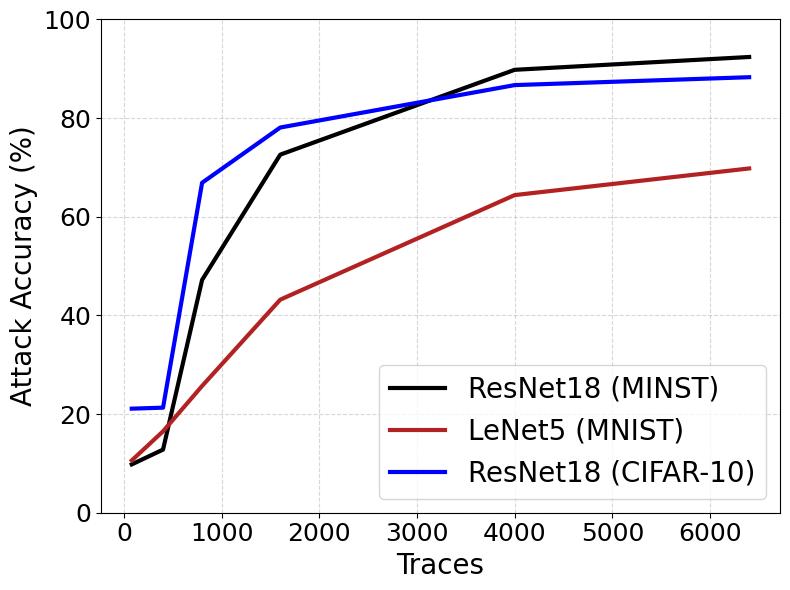}
    \caption{The relation between the attack accuracy of the attack model and the number of traces used in the attack.}
    \label{fig:numtracesacc}
\end{figure}

\ourmethod needs to profile the target device or a cloned device before mounting the attack. 
A greater number of profiling traces generally results in improved performance, but it also requires the adversary to access more resources, especially when the profiling device and the target device are the same.
We conduct experiments to investigate the relationship between the number of profiling traces and the attack effectiveness.
Figure~\ref{fig:numtracesacc} demonstrates the relationship between the amount of profiling traces and the attack accuracy for LeNet5 implementation on MNIST. 
It can be observed that 5\,000 profiling traces are sufficient to achieve an attack accuracy of 70\%.
Similar to techniques used in cryptographic research, averaging has proven effective for recovering inputs in neural network implementations. 
\ourmethod does not rely on per-sample averaging, even when multiple traces are collected for the same class. 
This highlights its efficiency of \ourmethod in utilizing raw traces for attribute prediction.

\subsection{Extraction with Misalignment}
\label{sec:misalignment}

\begin{figure}
    \centering
    \includegraphics[width=0.82\linewidth]{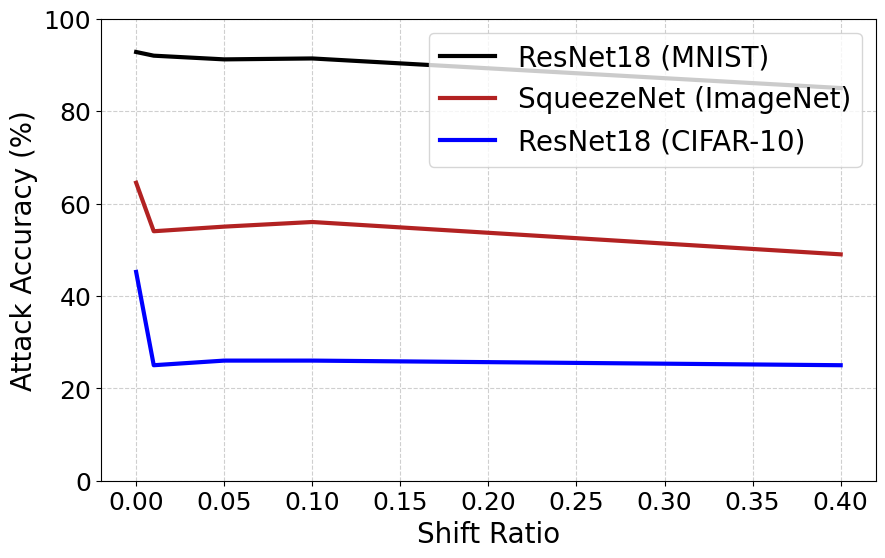}
    \caption{Comparison of attack robustness against trace misalignment, which is represented by the Shift Ratio.}
    \label{fig:miaalign}
\end{figure}

The adversary could collect the traces and conduct the alignment before the attack. 
In this section, we show that \ourmethod has the potential to attack with misaligned traces assuming possible measurement misalignment. 
Following the cryptographic side channel practices, we horizontally shift the test traces by a certain percentage of their original length (i.e., shift ratio) and report the attack accuracy of \ourmethod.
Figure~\ref{fig:miaalign} shows the prediction accuracy across varying amounts of horizontal shifts, where, for example, ratio 0.1 means that the attack traces can be randomly shifted in the range of  $[-0.1*\textnormal{length}, 0.1*\textnormal{length}]$ with zero padding. 
Even though there is a noticeable drop in attack accuracy, \ourmethod effectively handles trace misalignment and retains most of its effectiveness.

\subsection{Additional Analysis}
\label{sec:additional}

In this section, we conduct Test Vector Leakage Assessment (TVLA) on collected traces to analyze the possible operations that leak the input attribute.
We also provide a Grad-CAM analysis of the trained attack classifier for the same set of TVLA traces.

\subsubsection{t-SNE Visualization.}
\begin{figure}[t]
    \centering
    \includegraphics[width=0.6\linewidth]{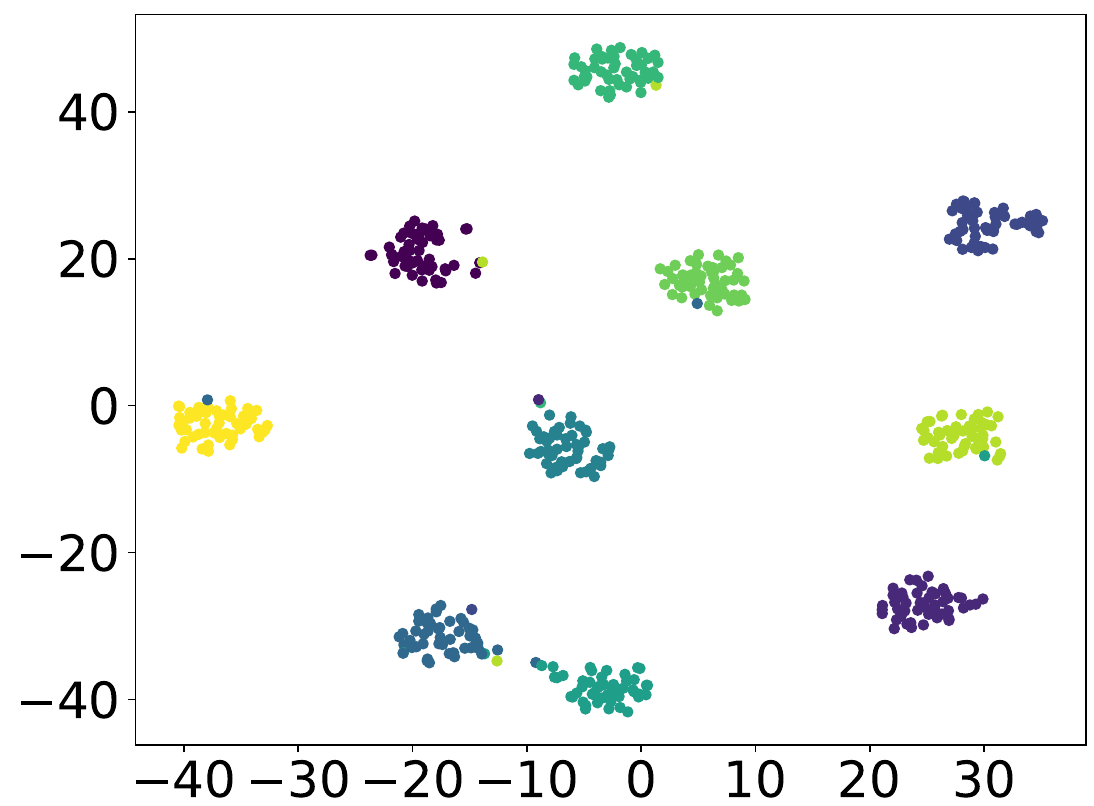}
    \caption{t-SNE visualization of 500 traces from 10 classes in feature space by \ourmethod with ResNet18 on MNIST}
    \label{fig:tsne}
\end{figure}

Figure~\ref{fig:tsne} present a t-SNE visualization to illustrate the trace distribution in feature space by \ourmethod with ResNet18 on MNIST.
Traces with different attributes are clearly separable in feature space of \ourmethod.

\subsubsection{Test Vector Leakage Assessment.}
Test Vector Leakage Assessment (TVLA) evaluates the presence of side-channel leakage in cryptographic implementations~\cite{schneider2015leakage}.
In particular, Welch's t-test determines if there is a significant difference between two sets of side-channel traces, where a significant difference suggests potential leakage of sensitive information. 
Regularly, TVLA compares measured traces from fixed versus random inputs to the implementation, after which it identifies possible vulnerabilities to side-channel attacks.

\begin{figure}[h]
    \centering
    \includegraphics[width=0.95\linewidth]{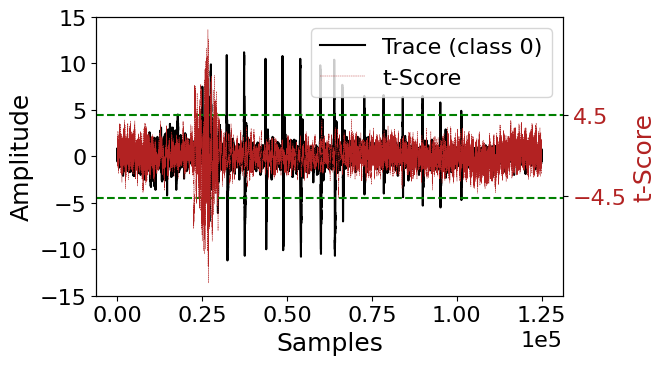}
    \caption{TVLA comparing class 0 traces with randomly selected traces from other classes.}
    \label{fig:tvla}
\end{figure}

We provide TVLA to determine if there is a significant difference between traces annotated with one class and traces with others. 
Figure~\ref{fig:tvla} demonstrates the TVLA results on EM traces collected from ZCU104 when running LeNet5 inference on the CIFAR-10 data set. 
A clear leakage can be observed near the first convolutional operation with a $|t| > 4.5$.
It suggests that the convolution operation may leak the class information.

\subsubsection{Grad-CAM Analysis.}

\begin{figure}[t]
    \centering
    \includegraphics[width=0.9\linewidth]{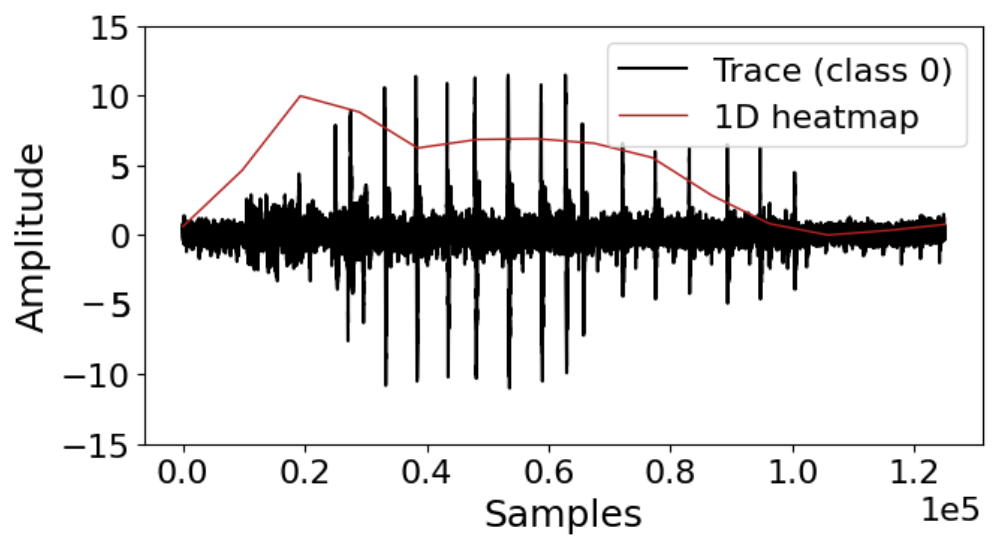}
    \caption{1D Grad-CAM of \ourmethod with CNN3 on MNIST.}
    \label{fig:grad}
\end{figure}

We also provide a post-attack analysis using 1-dimensional Grad-CAM~\cite{Selvaraju2017gradcam}.
After training, our \ourmethod model can predict the class from the EM trace. 
Given one sample trace with the correct prediction and the trained \ourmethod model on MNIST, we show the active part of the input trace calculated by Grad-CAM.
Interestingly, the Grad-CAM analysis (as shown in Figure \ref{fig:grad}) also indicates that the classification prediction is made mainly based on the traces representing the first convolution. 
Pre-attack analysis TVLA and post-attack analysis Grad-CAM reach the same conclusion, further verifying the effectiveness of \ourmethod.

\subsection{Other Discussions.}
\label{sec:otherdiscussions}

\noindent \textbf{Other DLSCA methods.} Two transformer-based DLSCA methods, 1D-Vision Transformer (1D-ViT) t and GPAM~\cite{bursztein2024generalized} are also considered in our attack.
Our experiment suggests that both 1D-ViT and GPAM do not outperform simple CNN architecture in our task.
We conjecture that transformer-based models are data-hungry, but our experimental setup has very limited amount of profiling data. 
We anticipate that more profiling data will increase the transformer-based DLSCA performance.

\noindent \textbf{Cross-session analysis.}
We evaluate \ourmethod in a cross-session scenario where the profiling physical setup differs from the attack setup, which better mimics the practical threat scenario. 
First, we profile the ZCU 104 running CNN3 on the MNIST dataset, and we train an attack CNN model that could extract the class attribute with an accuracy of 84.4\%. 
Then, during the attack, we apply the trained attack model to the EM traces collected during a new session where the measurement setup is re-mounted.
With a cross-session accuracy of 80.0\%, it indicates that \ourmethod generalizes across measurement sessions.

\noindent \textbf{Extraction on untrained CNN classifier.} We collect profiling traces on an untrained CNN3 classifier on MNIST.
Following the same attack pipeline, we observe that \ourmethod cannot predict attributes on EM traces collected from random CNN3 implementations. 
This experiment provides a controlled variable experiment, suggesting that the attribute leakage is from the neural network implementation itself instead of the input data, e.g., the AXI bus~\cite{Thu2023:YouOnlyGetOneShot}. 
This observation also confirms our finding that the convolution operation may leak as shown in Section~\ref{sec:additional}.

\subsection{Output Attribute Extraction}
\label{sec:output}

Previous attacks mainly focus on the input recovery attack due to the fact that input data represents the most privacy-sensitive information.
Model output, as part of the interactions, could also be privacy-sensitive~\cite{maji2024sparseleakynets}. 
For example, when a user aims to ask depression-related questions, the returned decision from either the search engine or the LLM agent would reflect sensitive information about the end user. 
By introducing an additional assumption about the adversary's capability in the threat model, concerning CNN and MLP implementations, we demonstrate that output attributes can also be recovered through the EM side-channel.

Following the original threat model, we assume that the adversary could access the target device and profile the target device, but at this time, the adversary could access the output label during profiling. 
Specifically, the adversary could compose a profiling set where the EM traces are also annotated with the model output classification label.
Table~\ref{tab:output} demonstrates that \ourmethod is also effective for output attribute extraction. 
As expected, when the input attribute accuracy is high, output attribute is also high since the ground truth values are quite similar. 
Note that, in several cases, output attribute accuracy is higher than input, suggesting that EM traces leaks more about output. 

\begin{table}[t]
\centering
\caption{Output attribute accuracy (\%) on MNIST, CIFAR-10, and ImageNet traces.}
\resizebox{0.8\linewidth}{!}{

\begin{tabular}{lccc}
\toprule
 Device&Implementation& MNIST & CIFAR-10\\
\midrule
\multirow{5}{*}{ZCU104}&MLP    &  44.9   &41.0  \\
&CNN3 & 83.5&-\\
&LeNet5  & 74.2&-\\
&SqueezeNet  &-&-\\
&ResNet18 &93.7&92.7\\
\bottomrule
\end{tabular}
}
\label{tab:output}
\end{table}

In this setup, our output attribute extraction attack is similar to the algorithm-level model extraction attack query-output pairs are collected to steal the target model~\cite{Tramer2016:StealingMLModels}. 
A difference is that we don't need the explicit input but the EM traces, which could make the adversary much stronger in edge applications. 
Consequently, we think that \ourmethod has a potential to improve algorithm-level model extraction when the adversary is in vicinity and could profile the target device. 

More importantly, we realize that in LLM applications, the output of the transformer model in the current step will be used as part of the model input of the next step.
It indicates that input recovery methods can be directly applied to extract LLM outputs seamlessly. 
We now provide a preliminary analysis in the next section.

\section{Edge LLM Experiments with RPi5.}
\label{sec:edgellm}

Edge LLM is an important application on edge~\cite{liu2024edge}, especially when applications require continuous and privacy-preserving inference~\cite{yu2024edge}.
Edge deployment of LLMs enables adversaries in the vicinity of the hardware device, making it an ideal target for physical side-channel attacks.
For the first time, we look at the side-channel leakages of LLMs with a focus on real-world devices.
In particular, we find EM side-channel leakages from the implementation of state-of-the-art edge LLM, \texttt{Qwen2-0.5B}\footnote{\url{https://huggingface.co/Qwen/Qwen2-0.5B-Instruct-GGUF}}~\cite{qwen2} with the help of LLaMa.cpp\footnote{\url{https://github.com/ggerganov/llama.cpp}} on the RPi5.
The performance of \texttt{Qwen2-0.5B} is documented in hugging face\footnote{\url{https://huggingface.co/Qwen/Qwen2-0.5B-Instruct}}.

\noindent \textbf{Experimental setup.}
As mentioned, Raspberry Pi 5 is a real-world target, given the inherent complexity of a full OS and black-box CPU level optimization. 
Our experiments are conducted on one RPi5 with 8GB RAM and quad-core 64-bit ARM Cortex-A76 CPU, where a large RAM is necessary to run the LLM inference on edge.

\begin{figure}[t]
    \centering
    \includegraphics[width=0.88\linewidth]{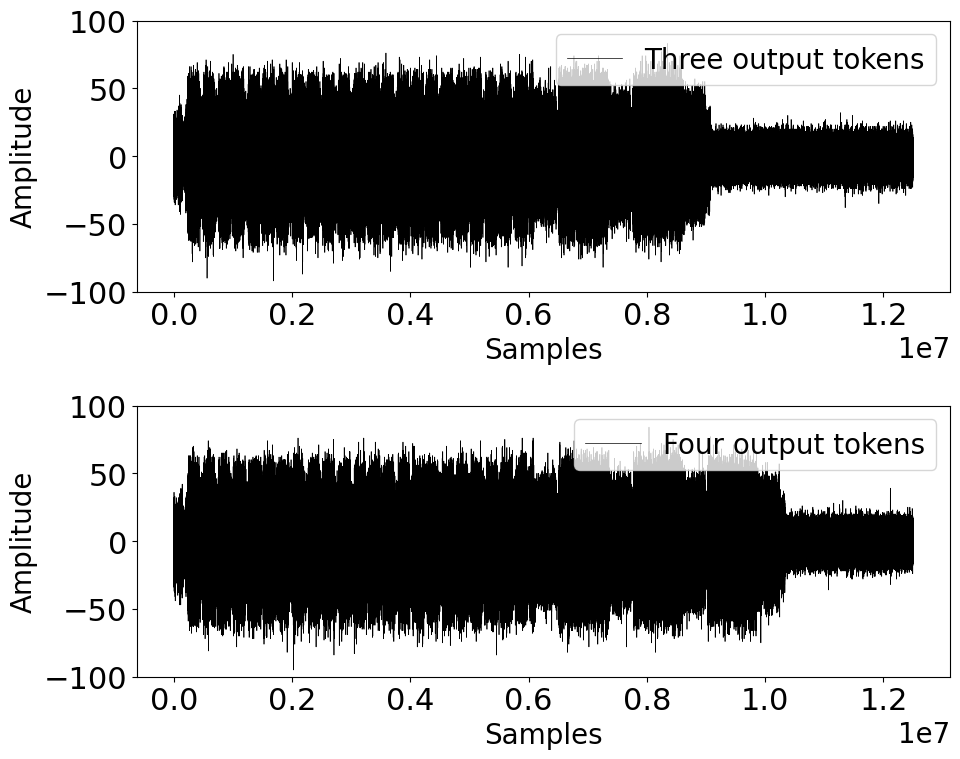}
    \caption{SEMA on RPi5 traces with different output tokens.}
    
    \label{fig:rpi5_trace_1_token}

\end{figure}

\noindent \textbf{SEMA of Qwen2-0.5B on RPi5.} Figure~\ref{fig:rpi5_trace_1_token} shows two EM traces corresponding to varying output lengths of the \texttt{Qwen2-0.5B}. 
It can be observed that an increase in the number of tokens included in the output results in an extended EM trace. 
The observed difference leaks the length of output tokens. 
Based on the regular design of the transformer-based LLM instruct model, the extra computational overhead is caused by additional tokens participating in the computations of the forward layer when previous tokens are cached.

\noindent \textbf{Distinguishing intermediate tokens.}
The token-by-token generation process of large language models (LLMs) presents a vulnerability that can be exploited through side-channel analysis. 
Built on the transformer architecture, LLMs iteratively feed intermediate outputs back into the model to produce the next output token.
This iterative mechanism can be leveraged by adversaries to perform side-channel attacks, as the same input recovery techniques can be applied to predict the outputs of previous iterations. 
In other words, adversaries who can measure EM emissions from the physical devices running the model can infer output attributes from the captured traces, effectively utilizing input recovery attacks.

\begin{figure}
    \centering
    \includegraphics[width=0.85\linewidth]{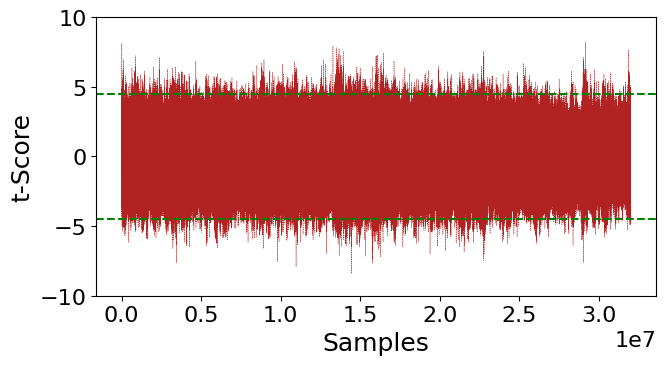}
    \caption{t-Test graph comparing the EM emissions of the Qwen2-0.5B-Instruct model processing two different intermediate output tokens.}
    \label{fig:LLMTVLAFIXFIX}
\end{figure}

We present an initial analysis of distinguishing between different intermediate tokens by comparing their corresponding EM traces following the same TVLA methodology~\cite{shukla2023whispering}.
We provide the same prompt to the Qwen2-0.5B-Instruct and aim to distinguish its first output token by examining the EM emissions.
Our assumption is that the EM traces corresponding to the second output token are evoked by taking the first output token as the input, where we can distinguish different intermediate tokens. 
This is an important observation since input recovery methodologies can help extract output tokens that may harm users' privacy. 
Figure~\ref{fig:LLMTVLAFIXFIX} shows the t-Test analysis aiming to distinguish input tokens `H' and `Sad' in `Happiness' and `Sadness'.
It can be observed that these two intermediate tokens are distinguishable in the EM trace with a $|t| > 4.5$.
Our experiments suggest that intermediate tokens (i.e., output tokens) can be distinguishable from EM traces, which has a potential to compromise user privacy.

\section{Outlook and Limitations}

\noindent \textbf{Identifying the operation that leaks.}
\ourmethod is effective in extracting attributes of user interactions with the system, and we also show that TVLA and Grad-CAM together with \ourmethod could provide useful information to identify the exact leak operation.
However, due to the black box nature of our physical target, it's hard to verify the effectiveness of \ourmethod in identifing and reasoning the exact operation that leaks.
Future work could leverage the \ourmethod methodology as a tool to assist secure neural network implementation. 

\noindent \textbf{Parallel work on the same device.}
\ourmethod is tested under conditions where the target device runs only a single inference task at a time. 
In practical scenarios, the situation could become more challenging when multiple inference tasks executing in parallel on the same physical device. 
While it is feasible to profile and launch attacks in such cases, it can be anticipated that the required task budget would be significantly higher. 
In this paper, we present a proof-of-concept to demonstrate the potential of \ourmethod.

\noindent \textbf{Measurement setups and computational resources.}
We provide proof-of-concept experiments to demonstrate that \ourmethod is implementation-agnostic and generalizable. 
However, some of our measurements may not be comprehensive in attacking real-world implementations. 
For example, the operating frequency of the RPi5 is adjusted to 300 MHz and the RPi3B to 600 MHz to match the bandwidth limitations of our oscilloscope. 
However, the memory capacity was still insufficient to capture the complete trace.
Because of the limited GPU memory when training on EM traces each has tens of millions of samples, our current DLSCA method has the potential to be optimized.
We show that \ourmethod has the potential to scale with more computational resources and powerful measurement equipment.

\section{Conclusion}
This paper investigates the use of electromagnetic side-channel analysis to extract private attributes from user interactions with neural network implementations on real-world devices.
We revisit previous input and output recovery attacks and suggest that the exact extraction of user interactions is not essential. 
To this end, we refine the threat model by shifting the focus from extracting exact interaction data to extracting private attributes associated with the interaction data.
Specifically, we propose a deep-learning-based profiling attack, \ourmethod, to extract private attributes. 
\ourmethod is trained on raw traces and predicts attributes by analyzing the traces collected during neural network inference.
Experimental results demonstrate the effectiveness of \ourmethod on different types of neural network implementations on real-world FPGA and Raspberry Pi.
Our attack is generic, implementation-agnostic, and easy to mount even without knowing the implementation details of the running devices.
We recommend incorporating our methodology as a standard evaluation step in future side-channel analysis on neural network implementations. 
Additionally, we suggest conducting further research to analyze vulnerabilities across various physical devices and implementations.

\bibliographystyle{plain}
\bibliography{ref}

\end{document}